\documentclass[12pt,preprint]{aastex}
\usepackage{amsmath}
\usepackage{graphicx}
\usepackage{color}
\usepackage{url}
\usepackage{CJKutf8}
\usepackage{ulem}

\slugcomment{DRAFT: accepted by Astronomical Journal}

\shorttitle{Phaethon activity 2016}
\shortauthors{Hui et al.}

\begin{document}

\title{Resurrection of (3200) Phaethon in 2016}
\author{
Man-To Hui 
\begin{CJK}{UTF8}{bsmi}
(許文韜)$^{1}$ 
\end{CJK}
and
Jing Li
\begin{CJK}{UTF8}{bsmi}
(李京)$^{1}$ 
\end{CJK}
}
\affil{$^1$Department of Earth, Planetary and Space Sciences,
UCLA, 
595 Charles Young Drive East, 
Los Angeles, CA 90095-1567\\
}
\email{pachacoti@ucla.edu}

\begin{abstract}

We present a study of the active asteroid (3200) Phaethon in the 2016 apparition using the \textit{STEREO} spacecraft and compare the results with data from the previous two perihelia in 2009 and 2012. Once again, Phaethon brightened by $\sim$2 mag soon after its perihelion passage, contradicting expectations from the phase function of a macroscopic monolithic body. Subsequently, a short antisolar tail of $\sim$0\degr.1 in length was formed within $\sim$1 day and quickly disappeared. No trail was seen. Our syndyne-synchrone analysis indicates that the tail was comprised of submicron to micron particles and can be approximated by a synchrone coinciding with the outburst. We estimate that the outburst has released a mass of $\sim$10$^{4}$--10$^{5}$ kg, comparable to the two mass ejections in 2009 and 2012, and that the average mass-loss rate is $\sim$0.1--1 kg s$^{-1}$. The forward-scattering effect hinted at low level activity of Phaethon prior to the outburst, which increased the effective cross section by merely $\lesssim$1 km$^2$. Without the forward-scattering enhancement, detecting such activity at side-scattering phase angles is very difficult. The forward-scattering effect also reinforces that the ejected dust grains rather than gas emissions were responsible for the activity of Phaethon. Despite Phaethon's reactivation, it is highly unlikely that the Geminid meteoroid stream can be sustained by similar perihelion mass-loss events. 
  
\end{abstract}


\section{\uppercase{Introduction}}

Asteroid (3200) Phaethon is dynamically associated with the Geminid meteoroid stream (Whipple 1983), along with several kilometre-sized asteroids collectively forming the Phaethon-Geminid complex (PGC; Ohtsuka et al. 2009). Unlike most meteoroid streams whose parents have been identified as cometary, members of the PGC are distinctly asteroidal objects, with an asteroid-like Tisserand parameter with respect to Jupiter of $T_\mathrm{J} \simeq 4.5$. Observations around perihelion in 2009 and 2012 at small solar elongations taken from the \textit{Solar and Terrestrial Relations Observatory} (\textit{STEREO}) have successfully revealed its activity, in terms of anomalous brightening (Jewitt \& Li 2010; Li \& Jewitt 2013) and subsequent development of a small tail (Jewitt et al. 2013). These studies argue that thermal fracture or desiccation cracking under very high surface temperature around perihelion are possible physical mechanisms for driving the activity in Phaethon, whereby small dust particles of $\sim$1 $\mu$m are released from the asteroid. Nonetheless, the observed mass loss is far from sufficient to sustain the Geminid stream (Li \& Jewitt 2013). Intriguingly, to date other similar small-perihelion asteroids show no evidence of mass loss (Jewitt 2013). Their faintness unfortunately hampers detection attempts of their activity in \textit{STEREO} images.

Currently the observations from \textit{STEREO} in 2009 and 2012 are the only two successful ones which show the activity in Phaethon.\footnote{Detection attempts for other apparitions have failed due to unfavourable viewing geometry. Notably Phaethon has been eluding the \textit{Solar and Heliospheric Observatory} (\textit{SOHO}) for two decades, because the spacecraft has an insufficient sensitivity and a narrower field-of-view.} If the mass-loss activity is driven by thermal fracture, the brightening and the formation of a tail should be recurrent in following apparitions under similar geometry between the Sun and Phaethon. In this paper, we analyse new \textit{STEREO} observations of Phaethon in 2016, and compare the measurements against those obtained from previous returns in 2009 and 2012. 

\section{\uppercase{Observations}}
\label{obs}
The observations of Phaethon were taken by one of the two Heliospheric Imagers (HI), HI-1, part of the Sun-Earth Connection Coronal and Heliospheric Investigation (SECCHI) package (Howard et al. 2008; Eyles et al. 2009) onboard the \textit{STEREO-A} spacecraft. The HI-1 camera covers regions along the ecliptic with the field centre offset from the Sun by 14\degr.0 in the ecliptic plane and has a square field-of-view of $\sim$$20\degr \times 20\degr$. Images taken by the HI-1 camera are usually $2 \times 2$ binned onboard to a dimension of $1024 \times 1024$ pixels, resulting in an angular size of 72\arcsec~for a binned pixel. Each HI-1 image is combined from 30 individual images with an exposure duration of 40 s onboard in order to remove cosmic rays and other transient objects in the field-of-view. In an ordinary mode, such an combined image is taken every 40 minutes. The effective optical response of the HI-1 camera is comprised of a major spectral bandpass of 630--730 nm (Eyles et al. 2009), with two leaks at $\sim$300--450 nm and $\sim$0.9--1.0 $\mu$m, respectively (Bewsher et al. 2010, see Fig. 6b). 

\begin{figure}
\begin{center}
\includegraphics[scale=0.543]{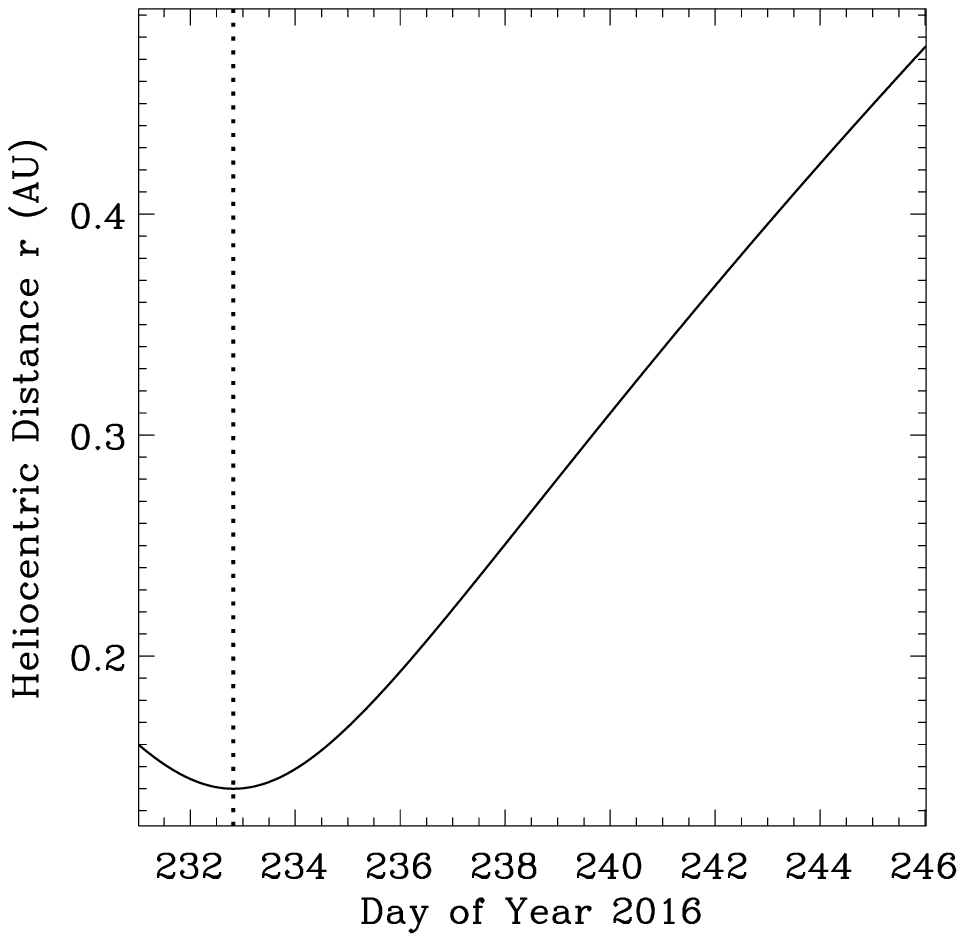}
\includegraphics[scale=0.543]{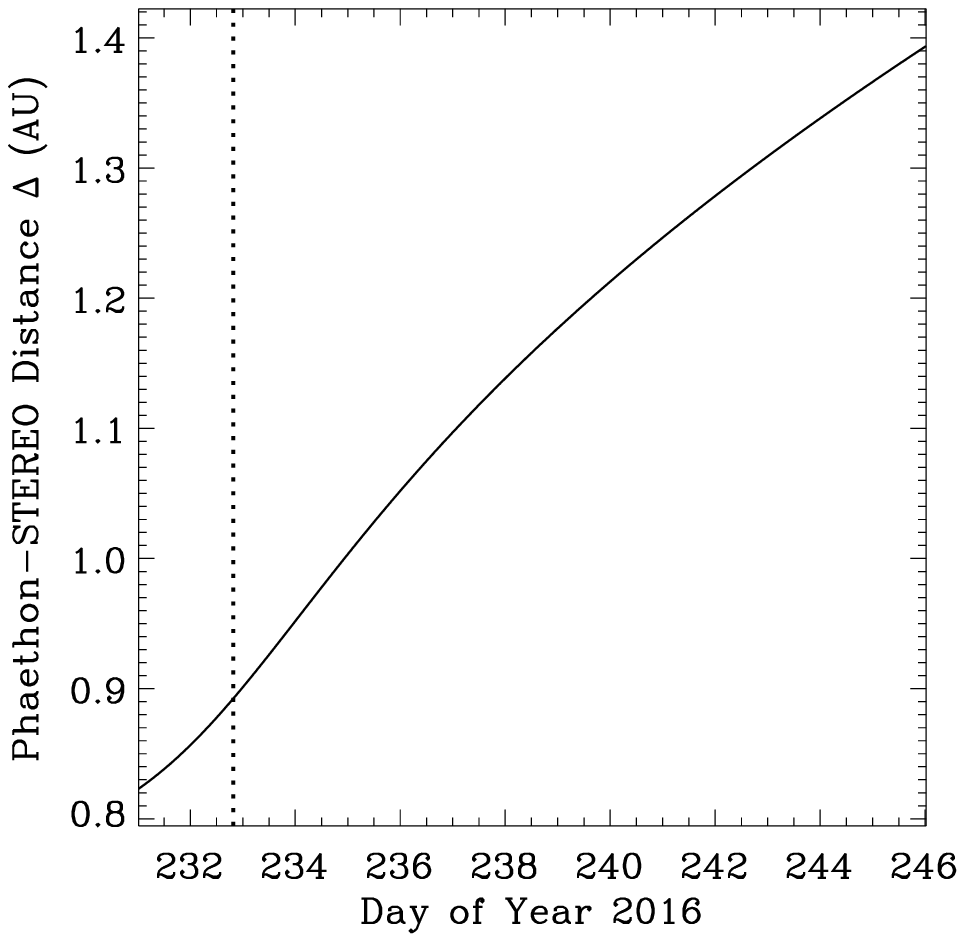}
\includegraphics[scale=0.543]{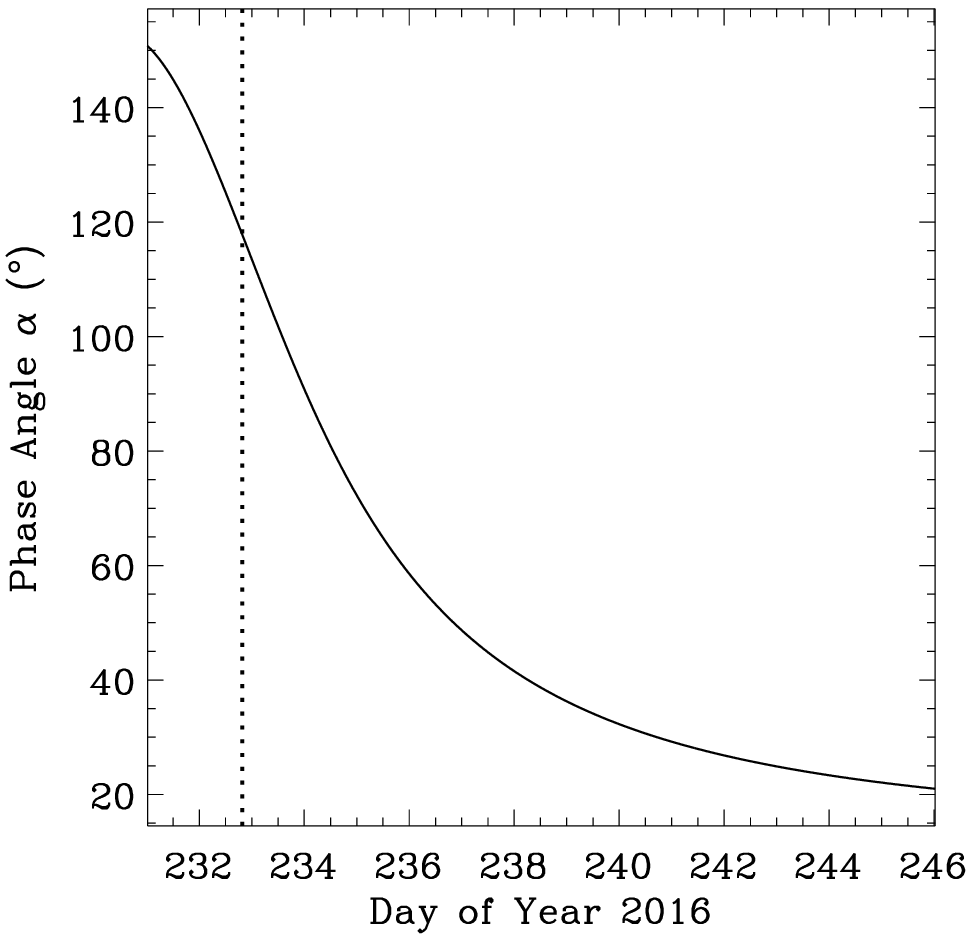}
\caption{
Observational geometry of (3200) Phaethon from the perspective of \textit{STEREO-A} during its perihelion return in 2016. The vertical dotted line in each panel marks the perihelion passage, i.e., UT 2015 August 19.82 (DOY = 232.82).
\label{fig_geo}
}
\end{center}
\end{figure}

\begin{figure}
\begin{center}
\includegraphics[angle=270,scale=1.0]{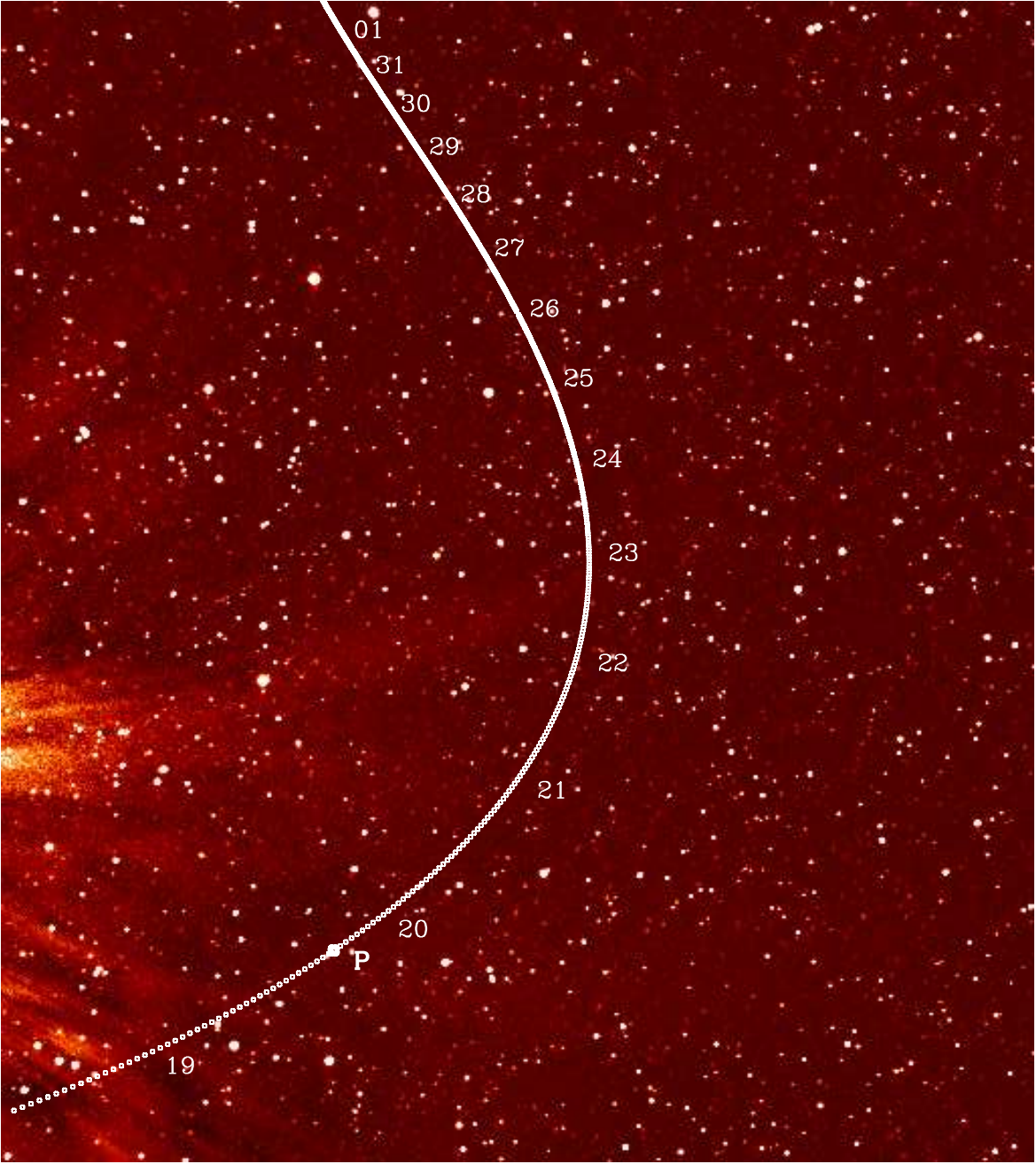}
\caption{
Trajectory of (3200) Phaethon in the field-of-view of the HI-1 camera of the \textit{STEREO-A} spacecraft. Numbers alongside the path label the day of August in 2016, with an exception that the number `01' near the upper edge is the day of September. Indicated by a thicker and larger circle as well as a bold-font letter `P' is the perihelion. Ecliptic north is up and east is left. The image has a field-of-view of $\sim$$10\degr.2 \times 11\degr.5$.
\label{fig_traj}
}
\end{center}
\end{figure}

Phaethon entered the field-of-view of the HI-1 camera around 2016 August 18.6 (DOY $\sim$ 231.6, day of year 2016) and was soon monitored to pass its perihelion passage on UT 2016 August 19.82 (DOY = 232.82). As it receded from the Sun, as well as moved away from the spacecraft, the apparent motion slowed down, gradually faded away, and became indistinguishable from background noise. Unlike the previous perihelion returns in 2009 and 2012 (Jewitt \& Li 2010; Li \& Jewitt 2013), Phaethon experienced the forward-scattering phase angle at the very beginning, and later days saw its phase angle gradually decrease (see Figure \ref{fig_geo}), which has provided a good chance to search for potential low activity pre-perihelion on Phaethon. The trajectory of Phaethon is illustrated in Figure \ref{fig_traj}.

\subsection{Photometry}
\label{phot}
We downloaded HI-1 camera level-0.5 images from the \textit{STEREO} Science Center\footnote{\url{https://stereodata.nascom.nasa.gov/index.shtml}}, which were later calibrated with bias and flat-field files to level-1 data by \textit{secchi\_prep} in the IDL-based SolarSoftWare (Freeland \& Handy 1998). The generated images are dominated by F-corona, which overwhelms field stars as well as Phaethon. To overcome this, we applied a technique similar to the one by Knight et al. (2010), but every background image was computed from minimum of neighbouring 14 images. Each of the level-1 images were then subtracted by the corresponding background image, yielding the final processed images that, by visual inspection, appear the most satisfactory, as the F-corona is largely removed, and no noticeable artefacts are seen. 

We employed the JPL HORIZONS\footnote{\url{http://ssd.jpl.nasa.gov/horizons.cgi}} to generate ephemerides for Phaethon as observed from \textit{STEREO-A}. Pixel coordinates of Phaethon in the HI-1 images were then computed following the method described in Thompson \& Wei (2010), which has been tested to be reliable enough, since the calculated pixel coordinates for several known field stars visible in the images, which were randomly chosen, are always superimposed by the corresponding real stars. Later we performed aperture photometry on the calculated pixel coordinates of Phaethon. By trial and error, a photometric aperture radius of 2-pixel was chosen as a compromise of maximising collection of the signal from Phaethon and minimising contamination from background sources. We do not expect a significant portion of the total flux beyond the chosen aperture, since the size is slightly over twofold the FWHM of the field stars in the HI-1 images (Bewsher et al. 2010), and also well encompasses the trailing of Phaethon (apparent motion $\lesssim$5\arcmin.6 hr$^{-1}$ during the observation, equivalent to a maximum trailing of $\sim$1.6-pixel in HI-1 images). The sky background was calculated as the median counts from an annulus circling around the photometric aperture, with an inner radius of 2-pixel and an outer radius of 6-pixel. 

\begin{figure}
\epsscale{1.0}
\begin{center}
\plotone{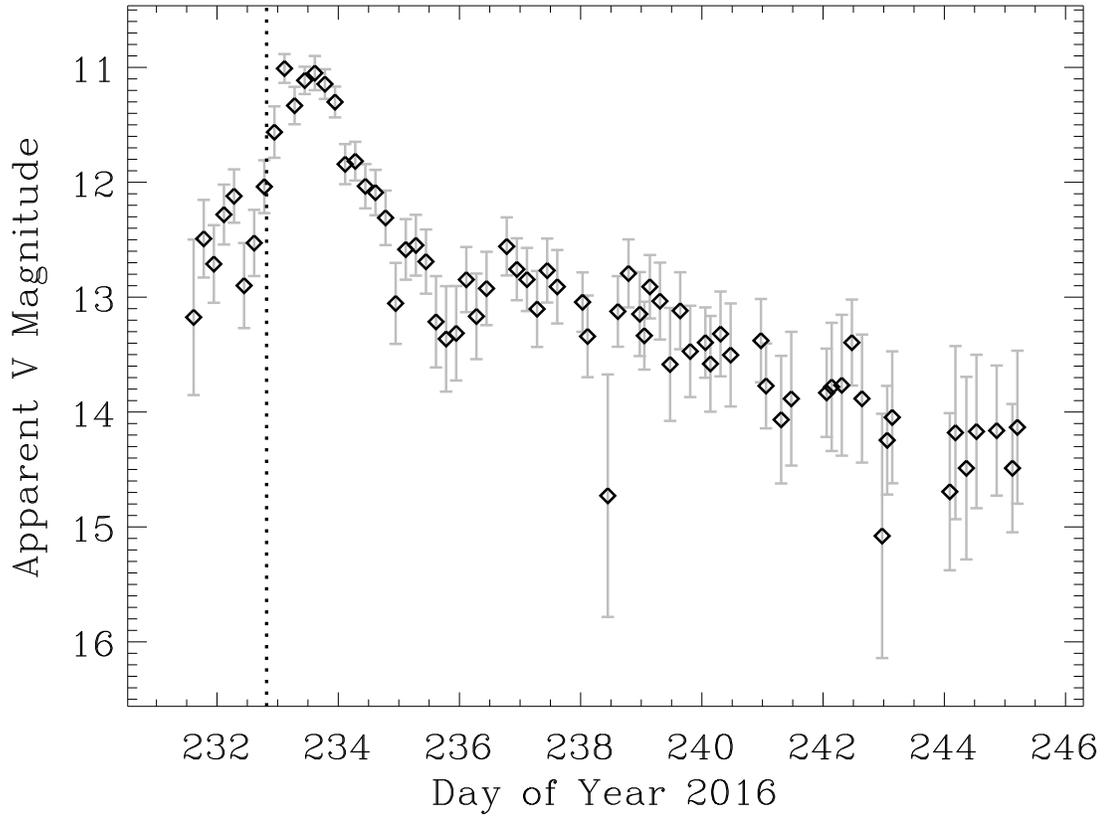}
\caption{
Apparent $V$-band magnitudes of (3200) Phaethon measured in HI-1 images as a function of time. The perihelion epoch of the asteroid is marked by a vertical dotted line. A surge in brightness around DOY = 233 (UT 2016 August 20) lasting for about three days is clearly seen.
\label{fig_lc}
} 
\end{center} 
\end{figure}

Our code performed photometry automatically on the HI-1 images. However, frequently field stars intruded into the photometric aperture, and worsens the measurements and leads to a huge scatter. We therefore proceeded to divide the whole image sequence into several groups, within which images were registered on Phaethon and rotated according to the antisolar angle (see Section \ref{morph}). For convenience we denote the number of images within each group as $\mathcal{N}$. By trial and error, we decided that every $\mathcal{N} = 6$ images were then median coadded to improve the signal-to-noise ratio of Phaethon. We re-performed photometry on these coadded images. The flux data were converted to HI-1 magnitudes, $m_\mathrm{HI}$, based on Bewsher et al. (2010 \& 2012). To obtain $V$-band magnitudes of Phaethon $m_V$, we applied the spectrum folding technique (e.g., Bewsher et al. 2010; Hui et al. 2015) to compute the magnitude difference between the HI-1 standard and $V$ band by

\begin{equation}
m_V - m_\mathrm{HI} = -2.5 \log \left[ \frac{\int F \left(\lambda\right) \mathfrak{T}_V \left( \lambda \right) \mathrm{d} \lambda }{\int F \left(\lambda\right) \mathfrak{T}_\mathrm{HI} \left( \lambda \right) \mathrm{d} \lambda} \cdot \frac{\int F_\ast \left(\lambda\right) \mathfrak{T}_\mathrm{HI} \left( \lambda \right) \mathrm{d} \lambda }{\int F_\ast \left(\lambda\right) \mathfrak{T}_V \left( \lambda \right) \mathrm{d} \lambda} \right]
\label{eq_dmag},
\end{equation}

\noindent where $F$ is the spectrum of Phaethon, $F_\ast$ is the spectrum of Vega, and $\mathfrak{T}$ is the effective transmissivity of some optical system, with the subscript $V$ and HI for labelling $V$ band and the HI-1 system respectively, and all are functions of wavelength $\lambda$.\footnote{The transmissivity of a $V$-band filter was downloaded from \url{http://spiff.rit.edu/classes/phys440/lectures/filters/filters.html}.} We utilised the spectra of Phaethon and Vega respectively from Licandro et al. (2007) and the stellar spectral flux library by Pickles (1998), and obtained $m_V - m_\mathrm{HI} = +0.38$ for Phaethon from Equation (\ref{eq_dmag}). The lightcurve of Phaethon in the $V$-band system is presented in Figure \ref{fig_lc}. We have tested different numbers of images for coaddition, such as every daily image sequence, i.e., $\mathcal{N} = 36$, and found that the resulting lightcurve does not alter within the uncertainty level. In this regard, it is noteworthy that our HI-1 photometry of Phaethon cannot be used to interprete lightcurve variations due to the spin, because the effective integration time of the coadded images is longer than its spin period ($\sim$3.6 hr; e.g., Hanu{\v s} et al. 2014), and also the uncertainty level is comparable to the reported amplitude ($\sim$0.1--0.4 mag; Ansdell et al. 2014).

\subsection{Morphology}
\label{morph}
\begin{figure}
\epsscale{1.0}
\begin{center}
\plotone{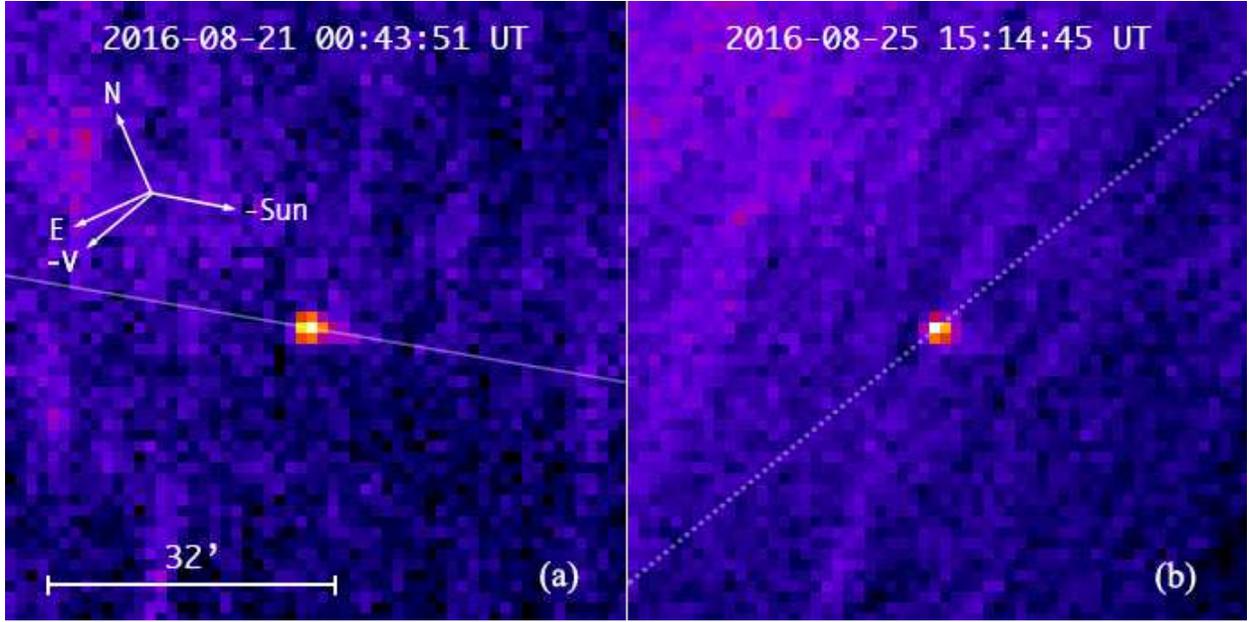}
\caption{
Appearance of (3200) Phaethon in the median combined HI1 images coadded from a daily sequence (a), and the whole image sequence (b), with alignment on the position angles of the antisolar direction ($\theta_\odot$) and the negative heliocentric velocity vector projected on the sky plane ($\theta_{\mathbf{V}}$), respectively. The equivalent mid-exposure epochs of the two images are written above. Orientation of the images is indicated by the compass, both referenced to the image taken on 2016 August 20 07:03:51 UT. The projected Sun-comet line (the narrow white line across the left panel) is drawn to better illustrate that the direction of the tail was antisolar, whilst the white dotted line across the right panel is the orientation of $\theta_{\mathbf{V}}$. A scale bar is shown as well, applicable to the two panels. The faint streaks are uncleaned trails of bright stars passing by the target.
\label{fig_3200}
} 
\end{center} 
\end{figure}

When blinking the HI-1 image sequence of Phaethon, we noticed that Phaethon seemingly had a faint tail, yet we could not be certain due to fluctuations in the sky background from image to image. In order to improve the signal-to-noise of Phaethon, we decided to follow the technique by Jewitt et al. (2013) and to stack the image sequence. We arbitrarily chose the HI-1 image taken on 2016 August 20 07:03:51 UT, when Phaethon was well within the field-of-view, as the reference image. Images were shifted to align on the calculated pixel coordinates of Phaethon, which were obtained in Section \ref{phot}. Since the position angle of the tail of Phaethon should change rapidly, rotation of the images is needed so that the signal-to-noise ratio of the tail can be improved. A real tail is confined between the antisolar angle and the negative heliocentric velocity vector, projected onto the plane of sky from some observer, denoted as $\theta_\odot$ and $\theta_{\mathbf{V}}$, respectively. A tail closer to the direction of $\theta_\odot$ is comprised of smaller particles, whereas closer to $\theta_{\mathbf{V}}$ is indicative of larger dust grains. The two quantities were both calculated by the JPL HORIZONS. Rotation of images about the calculated pixel coordinates of Phaethon respectively according to $\theta_\odot$ and $\theta_\mathbf{V}$ was then performed. The images were then median combined for further visual inspection. However, neither of the coadded images show a tail. We therefore decided to divide the whole image sequence into several groups. Within each group, the image sequence was median coadded into a single image. The first attempt with $\mathcal{N} = 36$, which is the number of daily images taken by the HI-1 camera, successfully reveals that Phaethon presented a short tail of $\sim$0\degr.1 (5-pixel) in length in coadded images taken starting from 2016 August 20 12:49 to August 21 12:38 UT, and were aligned with respect to $\theta_\odot$ (see Figure \ref{fig_3200}a). This feature was absent in images from other groups. The possibility of the tail being artificial has been ruled out, because none of the background stars in vicinity of Phaethon show similar structures in the HI-1 images, and the tail appears fainter or can be even washed out by rotating the images unphysically. We have tried several different smaller $\mathcal{N}$ as well, and all the images around the aforementioned period show a short tail pointing approximately to the antisolar direction, despite fainter. Therefore, we are confident that the Phaethon has shown mass-loss activity and presented a small tail comprised of small particles soon after its perihelion passage in 2016. The apparent length and position angle of the tail are both basically the same as the one observed around the previous perihelion passages in 2009 and 2012 (Jewitt et al. 2013).

While the detection of the near-antisolar tail is successful, we failed to recognise any tail-like structure around $\theta_{\mathbf{V}}$, or termed trail, however we selected $\mathcal{N}$ for image coaddition (see Figure \ref{fig_3200}b). In principle, the tail around $\theta_{\mathbf{V}}$ should have been best presented in the image coadded from the undivided image sequence, in that it consists of large dust grains, on which the solar radiation forces have less influence. The non-detection can be accounted by the following possibilities. First, the very low resolution of the HI-1 images may have hampered us from making such a detection, if the large dust grains did exist, because they are less susceptible to the solar radiation force than are the small grains, and a much longer time would be needed to cover the same distance (see Section \ref{sec_ds}). The second possibility is that the observed activity of Phaethon may have not released any large dust grains at all, or grains of these sizes are rare, and hence there was no sufficient effective cross section to scatter the sunlight.

\section{\uppercase{Discussion}}

\subsection{Phase Function}
\begin{figure}
\epsscale{1.0}
\begin{center}
\plotone{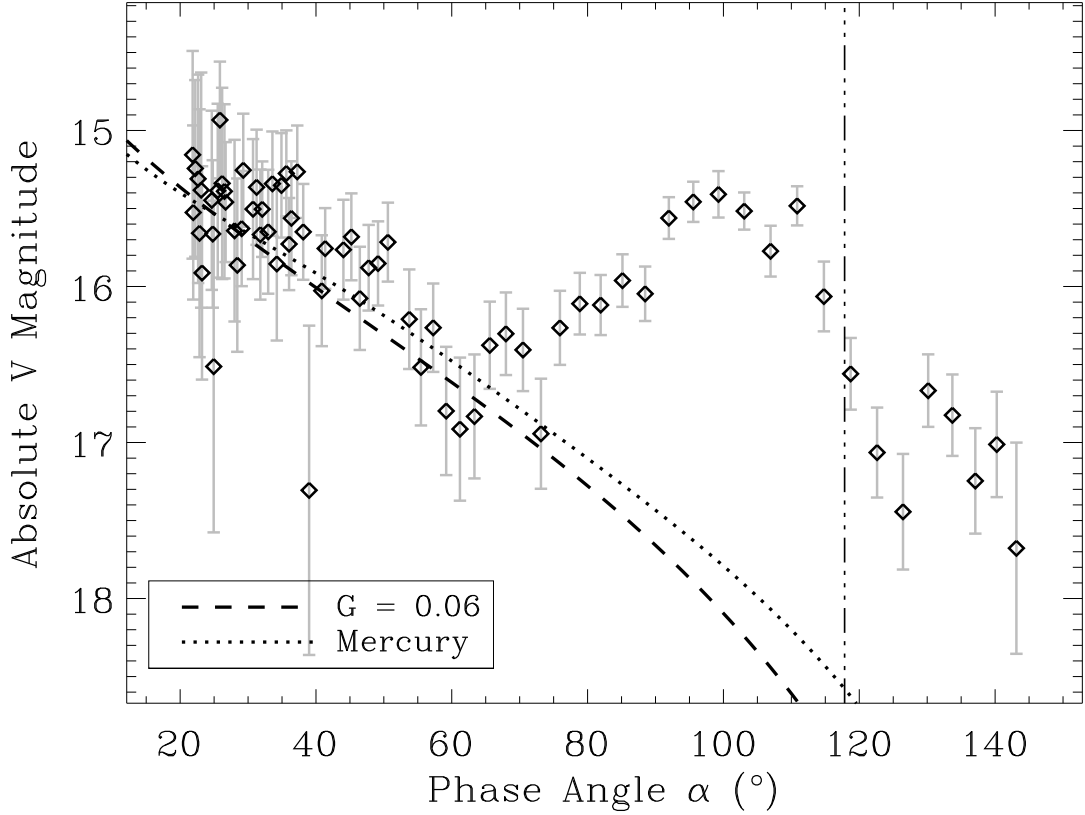}
\caption{
Absolute $V$-band magnitude data of (3200) Phaethon measured in HI-1 images as a function of phase angle $\alpha$. The brightening at $\alpha \gtrsim 80\degr$ contradicts what will be expected for a macroscopic monolithic object. The black dashed line is the phase function of the nucleus of Phaethon approximated by the HG formalism with G = 0.06 (Ansdell et al. 2014). Also plotted is a dotted line, which is the phase function of Mercury with a scaled $m_V \left(1, 1, 0\right)$ (Mallama et al. 2002). Phase angle at perihelion is marked by the vertical dashed-dotted line.
\label{fig_H_vs_phi}
} 
\end{center} 
\end{figure}

To correct for the variations in the heliocentric and Phaethon-\textit{STEREO} distances, denoted as $r$ and $\mathit{\Delta}$ respectively, we calculate the absolute magnitude $m_V \left( 1, 1, \alpha \right) $ as a function of phase angle $\alpha$, by assuming the inverse-square law, simply as

\begin{equation}
m_V \left( 1, 1, \alpha \right) = m_V \left(r, \mathit{\Delta}, \alpha \right) - 5 \log \left( r \mathit{\Delta} \right),
 \label{eq_H}
\end{equation}

\noindent where $m_V \left( r, \mathit{\Delta}, \alpha \right) $ is the $V$-band apparent magnitude, and both $r$ and $\mathit{\Delta}$ are expressed in AU. Figure \ref{fig_H_vs_phi} plots Equation (\ref{eq_H}) as a function of $\alpha$. Also plotted is the phase function in the HG formalism (Bowell et al. 1989) of Phaethon by Ansdell et al. (2014), which was best fit from observations made at phase angle $\alpha$ varying from 12\degr~to 83\degr. Inevitably a good range of the phase angle of Phaethon falls beyond $\alpha = 83\degr$ and extrapolation is needed. We therefore plot the phase function of Mercury by Mallama et al. (2002), which was obtained with a much wider range of $\alpha$, with a scaled absolute magnitude in the same figure as well. Obviously, the outburst in brightness was abnormal because it contradicts the phase function of a macroscopic monolithic object. It therefore likely indicates mass-loss activity on Phaethon.

Interestingly, Figure \ref{fig_H_vs_phi} suggests that Phaethon had some low level mass-loss activity even prior to the major outburst roughly one day pre-perihelion. Since the activity on Phaethon cannot be attributed by water-ice sublimation or prompt emission from forbidden transitions in atomic oxygen (Jewitt \& Li 2010; Li \& Jewitt 2013), the only remaining interpretation for the pre-perihelion brightness anomaly is the forward-scattering enhancement at large phase angle. Previous physical observations of Phaethon (e.g. Tedesco et al. 2004; Jewitt \& Hsieh 2006; Hanu{\v s} et al. 2016) have unambiguously unveiled the geometric albedo of Phaethon to be $p_V = 0.12$, and the effective diameter to be $D_\mathrm{N} \simeq 5$ km, or equivalently, the effective cross-section area of the nucleus as $C_\mathrm{N} = \pi D_{\mathrm{N}}^{2} / 4 \simeq 20$ km$^{2}$. Such small-scale activity increases the effective cross section by merely $\lesssim$1 km$^{2}$ (see Section \ref{sec_ml} and Figure \ref{fig_xs}), which corresponds to an increase in brightness by $\lesssim$0.05 mag during the side-scattering viewing geometry. This is at least a factor of two smaller than the magnitude amplitude due to the spin of Phaethon ($\gtrsim$0.1 mag; Ansdell et al. 2014), making the detection of activity difficult. On the other hand, the existence of the observed forward-scattering effect in turn strongly supports the argument in Li \& Jewitt (2013) that the brightness anomaly is due to the ejected dust grains from Phaethon, because gas emissions experience no forward-scattering enhancement whatsoever.

\subsection{Mass Loss}
\label{sec_ml}
\begin{figure}
\epsscale{1.0}
\begin{center}
\plotone{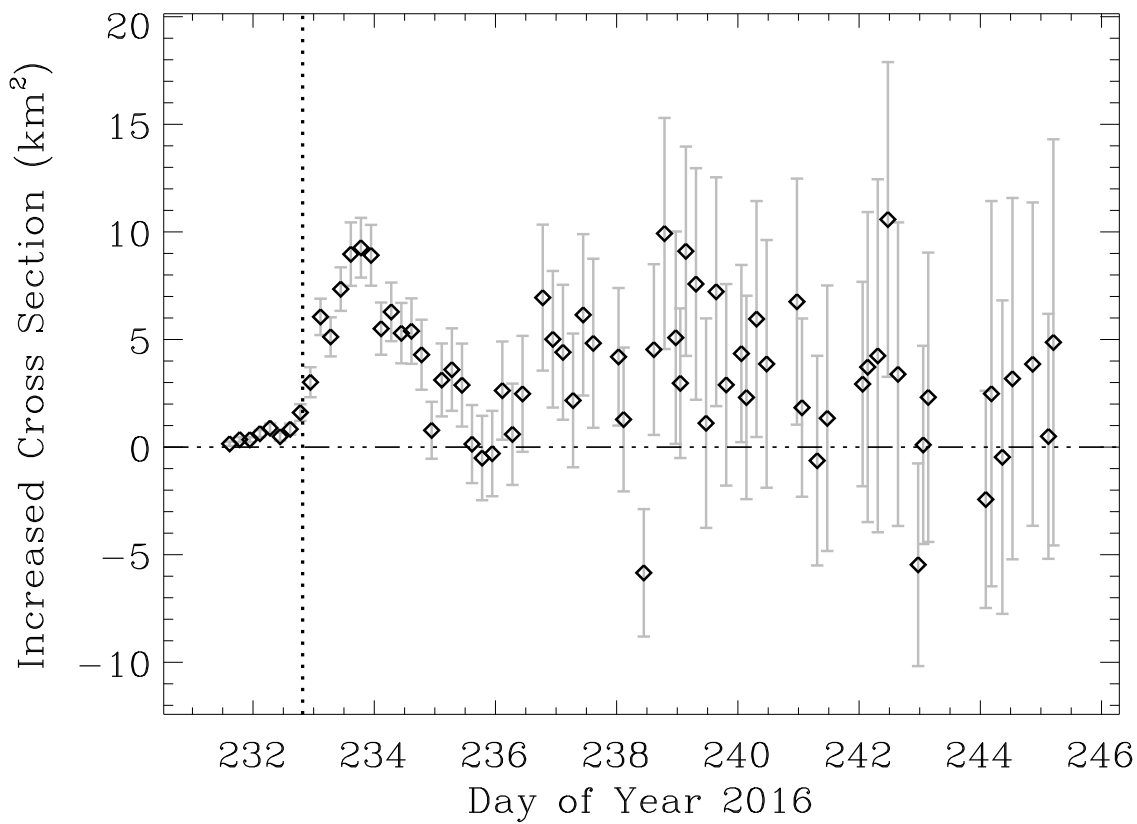}
\caption{
Increased effective cross section of (3200) Phaethon calculated from the HI-1 data as a function of time by Equation (\ref{eq_xs}). The perihelion epoch of the asteroid is labelled by a vertical dotted line, and the horizontal dashed-dotted line is where there is no effective cross-section increase. A surge in the effective cross section around DOY = 233 (UT 2016 August 20) is clearly visible. As the activity subsided, the effective cross section became indistinguishable from a bare nucleus, within the uncertainty level.
\label{fig_xs}
} 
\end{center} 
\end{figure}

The brighter absolute magnitude of Phaethon unambiguously indicates a larger effective cross section than that of the solid nucleus. We now estimate the increase in the effective cross section, $\Delta C$, based upon the HI-1 photometry data. Assuming that the cloud of dust grains released from the surface of Phaethon is optically thin, we obtain

\begin{equation}
\Delta C = \frac{\pi r_\oplus^{2}}{p_V \phi \left(\alpha\right)} 10^{-0.4 \left[ m_V \left( 1, 1, \alpha \right) - m_{\odot, V} \right]} - C_\mathrm{N} \frac{\phi_\mathrm{N} \left( \alpha \right)}{\phi \left( \alpha \right)},
\label{eq_xs}
\end{equation}

\noindent where $\phi$ is the dimensionless phase function of the dust grains, $\phi_\mathrm{N}$ is the phase function of the bare nucleus of Phaethon, $r_\oplus = 1$ AU expressed in km, and $m_{\odot, V} = -26.74$ is the apparent $V$-band magnitude of the Sun. We adopt the phase function by Marcus (2007) for the dust grains, and approximate $\phi_\mathrm{N} \left( \alpha \right)$ as the HG formalism with a slope parameter of G = 0.06 (Ansdell et al. 2014). Both phase functions are normalised at $\alpha = 0$, such that $\phi \left( 0 \right) = \phi_\mathrm{N} \left( 0 \right) = 1$. The uncertainty in $\Delta C$ is estimated by error propagation from the uncertainties in magnitude data, the geometric albedo and the effective cross section of the nucleus in Equation (\ref{eq_xs}). We plot the obtained $\Delta C$ against time in Figure \ref{fig_xs}.

The mass in spherical grains can be roughly approximated by

\begin{equation}
\Delta M = \frac{4 \rho \Delta C \bar{a}}{3}
\label{eq_dM}.
\end{equation}

\noindent Here $\bar{a}$ and $\rho$ are respectively the mean radius and the bulk density of the dust grains. The outburst about DOY $\sim234$ (UT 2016 August 21) is noticeably seen from Figure \ref{fig_xs}, which increased the effective cross section by $\sim$9 km$^2$. It is likely that Phaethon has gone through a second outburst starting from DOY $\sim 236$ (UT 2016 August 23), which was weaker than the earlier outburst and increased the effective cross section by $\sim$5 km$^{2}$. However, it is sensitive to the choice of the phase function for the bare nucleus of Phaethon. For instance, if Phaethon has a phase function more similar to that of Mercury than the approximation by the HG formalism, we then cannot find evidence for the second outburst, but are only left with the major outburst about DOY $\sim234$ (UT 2016 August 21). Therefore we focus on the major outburst only. With $\rho = 2.6$ g cm$^{-3}$ (Borovi\v{c}ka et al. 2010) and $\bar{a} \sim 1$ $\mu$m (see Section \ref{sec_ds}), Equation (\ref{eq_dM}) yields $\Delta M \sim 3 \times 10^{4}$ kg as a rough estimate for the mass loss of Phaethon. If the ejected grains obey a power-law dust size distribution similar to that of the Geminids, Equation (\ref{eq_dM}) can underestimate the mass loss by an order of magnitude. To include this uncertainty, we conclude that the mass loss of Phaethon around the perihelion is $\Delta M \sim 10^4$--10$^5$ kg, which is comparable to the mass loss during the previous two outbursts in 2009 and 2012 ($\sim$$3 \times 10^{5}$ kg; Jewitt et al. 2013). The increase in activity on Phaethon during the major outburst lasted for $\sim$1.2 days (see Figure \ref{fig_xs}). Thus the average mass-loss rate during this period is $\Delta M / \Delta t \sim 0.1$--1 kg s$^{-1}$, which is again comparable to the mass-loss rate in 2009 and 2012 ($\sim$3 kg s$^{-1}$; Jewitt et al. 2013), and also to some of the active asteroids such as 313P/Gibbs and 324P/La Sagra (c.f. Jewitt et al. 2015 and citations therein). 

The Geminid stream mass is $\sim10^{12}$--10$^{13}$ kg (Hughes \& McBride 1989; Jenniskens 1994). Were the Geminid meteoroid stream supplied by similar activity on Phaethon lasting for $\sim$1--2 days each orbit, the timescale for replenishing the Geminid stream has to be $\sim$1--100 Myr, which is much longer than the dispersion timescale of the Geminid stream ($\sim$1 kyr; Gustafson 1989). Therefore, it seems unlikely that the Geminid meteoroid stream can be sustained by the observed recurrent activity of such a small scale on Phaethon, as previously concluded (Jewitt et al. 2013).

The increased cross section started to decline after DOY = 233.7 (UT 2016 August 20.7), possibly because the resupply of dust grains from Phaethon failed to compensate those that quickly drifted out of the photometric aperture. Even worse, the uncertainties in the photometry observed especially after DOY $\sim 238$ (UT 2016 August 25) become so huge that we cannot unambiguously interpret the potential activity of Phaethon (see Figure \ref{fig_xs}). We thus conservatively conclude that there is no evidence about more activity on Phaethon after this date.

\subsection{Dust Size}
\label{sec_ds}
\begin{figure}
\epsscale{.7}
\begin{center}
\plotone{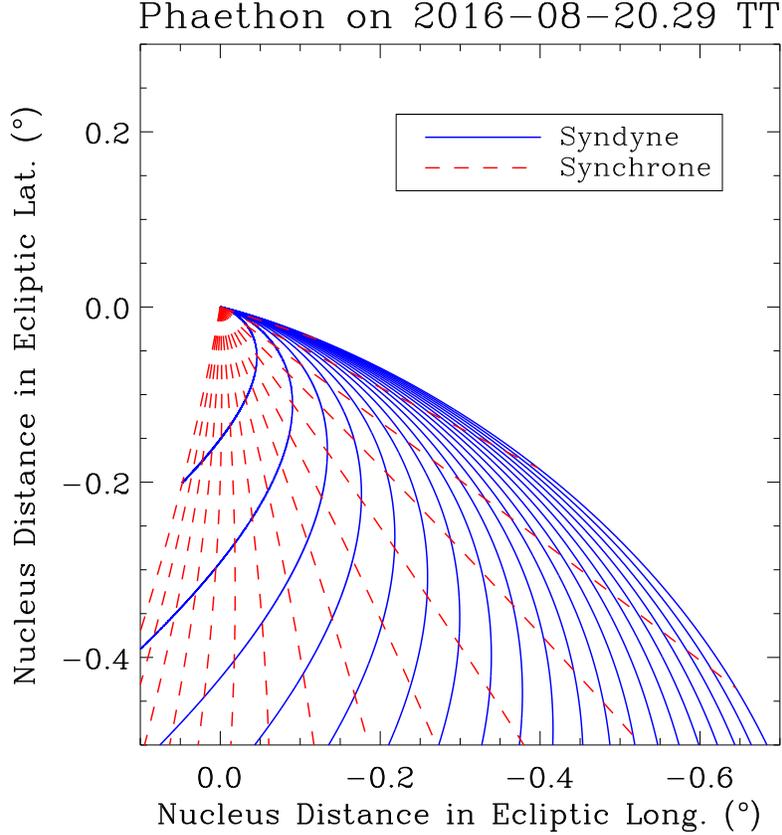}
\caption{
Syndyne-synchrone computation for (3200) Phaethon on 2016 August 20 07:03:51 UT, the referenced epoch. As indicated, the blue lines are the syndynes, and the synchrones are in dashed red. The synchrone lines correspond to ejections at fewer days prior to the epoch from left to right, with the leftmost 12 days, and a drawn interval of 0.8 days. The syndyne lines have decreased $\beta$ anticlockwise, with the rightmost $\beta = 0.5$ and a drawn interval of 0.025. Since we have no detection of the tail comprised of large dust grains, our concentration is on the syndyne-synchrone grid close to the antisolar direction, which remains basically unchanged during the HI-1 observation.
\label{fig_3200_fp}
} 
\end{center} 
\end{figure}

The apparent length of the tail was $\vartheta \sim 0\degr.1$, and it roughly spent $\tau \sim 1$ day growing and disappearing. Since the radiation acceleration of dust grains is related to the particle size, we can write

\begin{equation}
\bar{a} = \frac{3 \left( 1 + A_\mathrm{B} \right) S_\odot \tau^2}{8 c \rho r^{2} \mathit{\Delta} r_{\oplus} \vartheta}
\label{eq_a},
\end{equation}

\noindent where $A_\mathrm{B} = 0.04$ is the Bond albedo of Phaethon, converted from the geometric albedo based on Bowell et al. (1989), $S_\odot = 1361$ W m$^{-2}$ is the solar constant, $c = 3 \times 10^{8}$ m s$^{-1}$ is the speed of light, $r$ and $\mathit{\Delta}$ are both expressed in AU, and $\vartheta$ is expressed in radian. At the mid-exposure time (2016 August 21 00:43:51 UT), $r \simeq 0.15$ AU, $\mathit{\Delta} \simeq 0.95$ AU, Equation (\ref{eq_a}) yields $\bar{a} \simeq 1$ $\mu$m. This is consistent with the conclusion by Jewitt et al. (2013) drawn from the observations in 2009 and 2012. Particles of this size correspond to $\beta \sim 0.2$, the ratio between the radiation pressure force and the gravitational force due to the Sun. To verify this we compute a syndyne-synchrone diagram for Phaethon. A syndyne is the locus of particles of a common size, thereby the same $\beta$, released from the nucleus over a range of times, whereas a synchrone is the locus of particles released from the nucleus at the same time, but subjected to different $\beta$. The syndyne-synchrone computation assumes a zero initial ejection velocity for all of the particles (Finson \& Probstein 1968). Unfortunately because of the very low resolution of the HI-1 images, we are unable to firmly determine the $\beta$ value for the tail of Phaethon, as syndynes of large different $\beta$ tend to collapse together as approaching the antisolar angle (see Figure \ref{fig_3200_fp}). Nevertheless, by comparison against the observation, we can identify that the tail of Phaethon was comprised of submicron to micron particles with $\beta \lesssim 0.5$, and can be approximated better by a synchrone with a release time of $\lesssim$1 day before the observed epoch (i.e., around UT 2016 August 20), which roughly coincided with the start of the outburst in brightness. Note that there is a difference between the referenced epoch (2016 August 20 07:03:51 UT) we used to compute the syndyne-synchrone grid and the mid-exposure epoch. But this turns out to have little effect in our conclusion because the syndyne-synchrone lines close to the antisolar direction remain basically the same during the HI-1 observation. Therefore, we believe that Phaethon has undergone a brief mass-loss event soon post-perihelion, whereby small-sized particles were released from the nucleus. Similar events were observed in the HI-1 camera in the previous returns of Phaethon in 2009 and 2012 (Jewitt et al. 2013).

On the other hand, however, we cannot constrain the maximum dust grain size (or equivalently, the smallest $\beta$) from the HI-1 observations directly. Alternatively, we can still set a rough upper limit to the dust grain size from equilibrium between the solar radiation pressure and the gravitational accelerations due to Phaethon. Particles greater than the limit will fall back to the surface. We assume that Phaethon is a biaxial ellipsoid, with axis radii $R_1 > R_2 = R_3$. With some algebra we obtain

\begin{equation}
a_{\max} =  \frac{9 \left( 1 + A_\mathrm{B} \right) S_\odot f^{3/2} }{8 \pi c G \rho^2 r^2 D_\mathrm{N}},
 \label{eq_amax}
\end{equation}

\noindent where $f = R_1 / R_3$, which is constrained from the rotational lightcurve of Phaethon as $f = 1.45$ (Jewitt \& Hsieh 2006), $G = 6.67 \times 10^{-11}$ m$^{3}$ kg$^{-1}$ s$^{-2}$ is the gravitational constant, and $r$ is expressed in AU. Inserting values into Equation (\ref{eq_amax}) yields $a_{\max} \simeq 65$ $\mu$m for Phaethon around the perihelion. Since the Geminid stream contains a significant portion of submillimetre sized meteoroids (Borovi{\v c}ka et al. 2010), we argue that, besides the timescale problem (see Section \ref{sec_ml}), meteoroid-sized dust grains are not favoured by perihelion mass-loss activity of the type we observed. Given the crude estimate where the direction of the radiation pressure acceleration relative to the nucleus centre is ignored, and dust particles are ejected up with some initial speed and can reach some height above the surface, thereby a weaker gravitational acceleration, we cannot completely rule out the possibility that a small fraction of the Geminid meteoroids might have come from perihelion outburst events. Nevertheless, it reinforces the argument that the Geminid meteoroid stream cannot be produced by the observed near-Sun activity of Phaethon alone.

\section{\uppercase{Summary}}
We present the analysis of Phaethon during the perihelion in 2016 observed by the HI-1 camera onboard the \textit{STEREO-A} spacecraft. Key conclusions of the results are summarised as follows.

\begin{enumerate}
\item Phaethon has experienced a major outburst in brightness starting around perihelion, at DOY $\sim 233$ (UT 2016 August 20), when it brightened by $\sim$2 mag. The abnormal brightening contradicts the prediction from a macroscopic monolithic object.

\item The outburst in brightness was due to a mass-loss event on Phaethon, which increased the effective cross section to scatter more sunlight. Thanks to the event a short antisolar tail comprised of particles with $\beta \lesssim 0.5$, i.e., submicron to micron in radius, was formed within $\sim$1 day.

\item We estimate that Phaethon has released $\Delta M \sim 10^{4}$--10$^{5}$ kg during the current mass-loss event, which is similar to the previous two events observed in 2009 and 2012 around its perihelion. The average mass-loss rate during the major outburst was $\sim$0.1--1 kg s$^{-1}$, comparable to some of the active asteroids.

\item The forward-scattering enhancement prior to the major outburst indicates that Phaethon had low level activity roughly within one day pre-perihelion, where the effective cross section was increased by $\lesssim$1 km$^{2}$. Without the help of the forward-scattering effect, it is challenging to identify such activity, in comparison to the lightcurve amplitude due to its spin.

\item The existence of the observed forward-scattering effect shows that it is ejected dust grains that are responsible for the abnormal brightening of Phaethon, rather than gas emissions.

\item It is highly unlikely that the Geminid meteoroid stream can be sustained by mass loss events of such small scale within the dispersion timescale of $\sim$1 kyr.



\end{enumerate}

\acknowledgements
We thank amateur astronomers who reported the abnormal activity of Phaethon in the Yahoo mailing list, where we were alerted, and David Jewitt, Henry Hsieh, and Paul Wiegert for valuable discussions. {In particular,} comments from our referee Matthew Knight and David Jewitt on the manuscript have helped us greatly. MTH appreciate the discussions with Matthew Knight on the effective transmissivity of the HI-1 camera several years back. This study has inclusively utilised products from the Heliospheric Imager (HI), the instrument developed by a collaboration including the US Naval Research Laboratory (NRL), Washington, DC, USA, the University of Birmingham and the Rutherford Appleton Laboratory, both in the UK, and the Centre Spatial de Li{\`e}ge (CSL), Begium. The \textit{STEREO}/SECCHI project is an international consortium. This work is funded by a grant from NASA to David Jewitt.

\end{document}